\documentclass[]{aa}
\usepackage{graphicx}
\usepackage{txfonts}
\usepackage{natbib}
\usepackage{balance}
\usepackage[figuresright]{rotating}
\usepackage[a4paper,breaklinks]{hyperref}

\idline{1}{1}
\begin{document}
\def\teff{$\mathrm T_{eff }$ }
\def\logg {log\,g}
\def\lambo{$\lambda$ Boo }
\def\vsini {$v\,\sin i$ }
\def\kms {$\mathrm{km\, s^{-1}}$ }
\newcommand{\pun}[1]{\,#1}
\newcommand{\loggf}{\ensuremath{\log\,gf}}
\newcommand{\mlp}{\ensuremath{\alpha_{\mathrm{MLT}}}}
\newcommand{\LHD}{{\sf LHD}}
\newcommand{\xx}{\ensuremath{\mathrm{1D}_{\mathrm{LHD}}}}
\newcommand{\cobold}{{\sf CO$^5$BOLD}}
\newcommand{\mD}{\ensuremath{\left\langle\mathrm{3D}\right\rangle}}
\newcommand{\linfor}{{\sf Linfor3D}}

\title{The solar photospheric nitrogen abundance}
\subtitle{Analysis of atomic transitions with 3D and 1D model atmospheres}

\author{
E.~Caffau     \inst{1}
\and
E.~Maiorca    \inst{2}
\and
P.~Bonifacio  \inst{3,1,4}
\and
R.~Faraggiana \inst{5}
\and
M.~Steffen    \inst{6}
\and
H.-G.~Ludwig  \inst{3,1}
\and
I.~Kamp       \inst{7}
\and
M.~Busso      \inst{2,8}
}

\institute{ 
GEPI, Observatoire de Paris, CNRS, Universit\'e Paris Diderot; 92195
Meudon Cedex, France
\and
Department of Physics, University of Perugia, via Pascoli, Perugia, I-06123, Italy
\and
CIFIST Marie Curie Excellence Team
\and
Istituto Nazionale di Astrofisica,
Osservatorio Astronomico di Trieste,  Via Tiepolo 11,
I-34143 Trieste, Italy
\and
Dipartimento di Astronomia, Universit\`a 
degli Studi di Trieste, 
via G.B. Tiepolo 11, 34143 Trieste, Italy
\and
Astrophysikalisches Institut Potsdam, An der Sternwarte 16, D-14482 Potsdam, Germany
\and
Kapteyn Astronomical Institute, Postbus 800, 9700 AV Groningen
\and
Istituto Nazionale di Fisica Nucleare, section of Perugia, via Pascoli, Perugia, I-06123, Italy
}

\mail{}
\authorrunning{Caffau et al.}
\titlerunning{The Solar photospheric nitrogen abundance}

\date{Received 27 August 2008 / Accepted 30 January 2009}
\abstract
{In recent years, the solar chemical abundances have been studied in considerable detail because of 
discrepant values of solar metallicity inferred from different indicators, i.e., 
on the one hand, the ``sub-solar'' photospheric abundances resulting from 
spectroscopic chemical composition analyses with the aid of 3D hydrodynamical 
models of the solar atmosphere, and, on the other hand, the high metallicity inferred by helioseismology.
}
{After investigating the solar oxygen abundance using
a \cobold\ 3D hydrodynamical solar model in previous work, we 
undertake a similar approach studying the solar abundance of nitrogen, 
since this element accounts for a significant fraction of the overall solar
metallicity, $Z$.}
{We used a selection of atomic spectral lines to determine the solar
nitrogen abundance, relying mainly on equivalent width 
measurements in the literature.
We investigate the influence on the abundance analysis,
of both deviations from local thermodynamic equilibrium 
(``NLTE effects'') and photospheric inhomogeneities (``granulation effects'').}
{
We recommend use of a solar nitrogen abundance of
A(N)=$7.86\pm 0.12$\thanks{A(N)$\equiv \log n(\rm N) - \log n(\rm H) +12$},
whose error bar reflects the line-to-line scatter.
}
{The solar metallicity implied by the
{\cobold}-based nitrogen and oxygen abundances is in the
range $ 0.0145\le Z \le 0.0167$. 
This result is a step towards reconciling 
photospheric abundances with helioseismic constraints on $Z$.
Our most suitable estimates are $Z=0.0156$ and $Z/X=0.0213$.}

\keywords{Sun: abundances -- Stars: abundances -- Hydrodynamics -- Line: formation}            
\maketitle

\section{Introduction}

The knowledge of the precise nitrogen abundance in the solar photosphere 
is important because N accounts for about 6\% (by mass) of the solar 
metallicity, and hence is one of the most important contributors to $Z$
after O (53\%), C (23\%), and Ne (14\%). Amongst others,
the value of $Z$ has a crucial impact on stellar structure calculations. 
The abundance of C, N, and O in the solar system can be 
determined only from the solar spectrum. 
Meteoritic samples are un-representative, because C, N, and O 
are volatile elements. With condensation temperatures of only a few hundred 
Kelvin, they condense only partially so that their abundances in pristine 
meteorites (e.g., carbonaceous chondrites) do not provide valid estimates 
of the average solar system composition. The main molecule 
formed by C and O, carbon monoxide (CO), is also extremely stable. As a consequence,
all C will be locked into CO if the original material is O-rich, and will
be unable to form solid compounds. Only the remaining  C may condense into grains, 
and in the resulting meteoritic sample carbon will be underrepresented. 
The reverse is true if the environment is C-rich: all oxygen remains in the
gas phase and the remaining C may form grains. Therefore, the global elemental 
C/O ratio cannot be inferred from meteorites; although, 
isotopic ratios can be measured with high accuracy.
Similar problems affect nitrogen in an even more severe manner.
As a consequence, the meteoritic C, N, and O abundances are lower than the 
solar photospheric ones by large factors: 2-3 for oxygen, about 10 for 
carbon, and more than 40 for nitrogen \citep[see e.g.,][]{palme05}.
Analyses of observations of the solar corona also do not provide reliable abundance
measurements of these
elements. Several processes affect the solar corona composition,
partially changing the abundance of some elements 
\citep{ag89,gs00}.
The solar nitrogen abundance can be derived reliably only from the photospheric spectrum.

Besides its importance in determining the
solar metallicity, nitrogen is an interesting element
from a nucleosynthetic point of view.
Its main production channel is the CNO cycle,
in which it is produced at the expense of C and O.
The main astrophysical N production sites remain unclear;
AGB stars are good candidates, in addition to rotating massive stars.
Nitrogen is measured in many different objects: hot stars, cool stars,
Galactic and extragalactic \ion{H}{ii} regions, and Damped Ly$\alpha$
systems at high redshift. For all of these studies, a good solar
reference value is of fundamental importance.

The measurements of heavy element abundances inferred from photospheric solar spectra 
should agree with results from helioseismology.
Helioseismology can provide a measurement of the 
solar metallicity in essentially three ways:
i) from the depth of the convection zone:
this depends sensitively on the opacity at the base of the
convection zone, which in turn is a function of the 
abundance of heavy elements \citep{BA97};
ii) using information from the core: the small-frequency 
spacings of low-degree modes and their separation ratios
are sensitive to the mean molecular weight in the core, which can
be related to the metallicity of the outer layers \citep{basu07}; 
iii) from the sound speed gradient in the ionisation zone:
the depth profile of this quantity can be inferred from helioseismic
inversions, and comparison with the results of theoretical solar 
structure models of different metallicity allows us to place tight
constraints on $Z$
\citep{antia}.
A comprehensive review on helioseismology and solar
abundances is given by \citet{basu} which the reader is referred to 
for further details. It is remarkable that all helioseismic
methods provide results that are fairly consistent 
with each other, which infer a higher metallicity 
than currently deduced from the analysis of the photospheric spectrum.

In this paper, we determine the photospheric nitrogen abundance
by using data for atomic lines and the latest \cobold\ solar model.
One of our aims is to combine this result with our other 
solar abundance determinations to obtain a measure of the
solar $Z$ based on the \cobold\ solar model, and to compare
this with the $Z$ values inferred by helioseismology.


\section{Nitrogen abundance indicators}

\subsection{Forbidden lines}
The strongest forbidden [\ion{N}{i}] lines within the ground
configuration are close to 1040.0\pun{nm} 
and belong to the multiplet $^2$D$^0$-$^2$P$^0$. 
\citet{Houziaux} claimed to detect these
lines in the solar spectrum, although studies 
by \citet{LS67} and \citet{Swensson} could not confirm these detections.
\citet{Swensson} also demonstrated convincingly
that the predicted equivalent widths are too weak 
to be detectable in the Jungfraujoch Atlas.

\subsection{Permitted lines}
\begin{figure}
\resizebox{\hsize}{!}{\includegraphics[clip=true,angle=0]{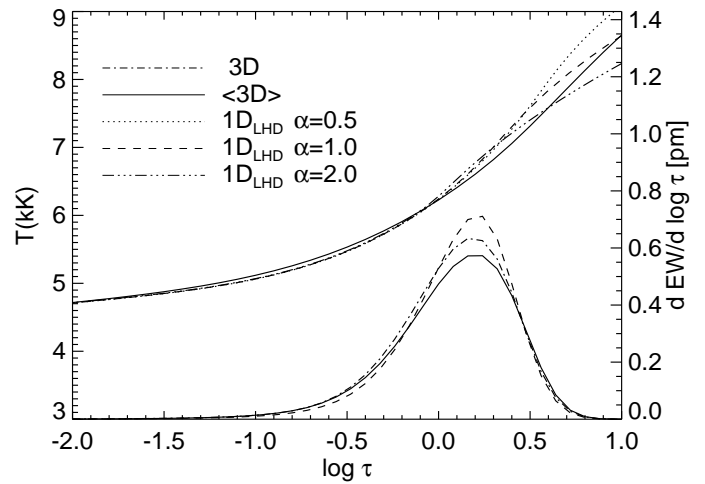}}
\caption{The temporal and horizontal average of the temperature profile of the 3D
  model (solid line) and the temperature profile of \xx\ models of three
  values of \mlp, are shown as a function of (Rosseland) optical depth.
  In addition, equivalent width contribution functions (lower, roughly 
  Gaussian-shaped curves) for the 746.8\pun{nm} line at disc-centre are 
  plotted as a function of monochromatic continuum optical depth.}
\label{ttau_conft}
\end{figure}

The permitted \ion{N}{i} lines of low excitation potential 
occur at UV wavelengths, mostly in the 80-120\,nm  range. 
The atomic \ion{N}{i} lines measurable in the solar spectrum 
are found in the visual and near IR range, and all have a high excitation potential 
(E$_{\rm low} >$ 10.3~eV), implying that these lines are formed in the 
deep layers of the photosphere. They are therefore hardly
affected by departures from LTE 
\citep[$\Delta\,$A(N)$\,\sim\,$$-0.05$~dex,][]{inga96}.
In the classical analysis that is based on 1D model atmospheres, these lines
have the disadvantage of being very sensitive to the temperature
structure of the deep layers, which, for theoretical
models, depends on the treatment of convection and,
most significantly, on the choice of the mixing-length
parameter (see Fig.\,\ref{ttau_conft}).
In principle, these difficulties are overcome 
by modern 3D hydrodynamical simulations, which provide a physically
consistent description of convection from first principles.

\subsection{Molecular lines}
\citet{grevesse} illustrated that atomic and molecular lines
react in opposite directions to temperature perturbations, so that the 
agreement between the atomic and molecular lines provides support for 
the choice of solar model. In their molecular analysis,
they and other authors have relied on the use of NH and CN molecular 
lines. In this paper, we derive the solar N abundance from only the atomic
lines and defer the study of molecular lines to the future.
On the one hand the abundance of carbon has not yet been estimated with a
\cobold\ model, and on the other hand, synthesising the NH A-X band,
used e.g. by \citet{lambert78} in the
solar spectrum, would require accounting for a large number of lines, both
NH lines and blending atomic lines, and the state-of the art version of
\linfor\ is unable to handle hundreds of lines.
Potential deviations from LTE for the molecular species is another
issue that hampers the use of molecular lines for accurate abundance analysis.
For these reasons, we postpone the study of molecular lines harbouring N.


\subsection{Line selection}
The first selection of lines was completed by extracting all the \ion{N}{i} lines
from the NIST Atomic Spectra Database \citep[NIST ASD][]{Ralchenko} in the
range 400-1200\,nm. These lines were then checked by visual inspection in
the solar atlases of \citet{delbouille} and \citet{delbouilleir}; the
identifications were achieved by using the Utrecht spectrum identifications
\citep{moore66} for the 400-877\,nm range, and those by \citet{Swensson70} for 
the 900-1200\,nm range. According to the NIST ASD, no \ion{N}{i} line is present in the
877-900\,nm range, which is not covered by these identification lists.  We also
performed a check of the relative line intensities with the Moore
Multiplet Table\,\citep{Moore}. The strongest and least blended lines 
are given in Table\,\ref{atomicdata} with the \loggf\ values, and their
qualities are taken from the NIST ASD.

\begin{table}
\caption{Selection of permitted \ion{N}{i} lines in the visual 
and near IR spectral range}
\label{atomicdata}
\begin{center}
\begin{tabular}{rcclrr}
\noalign{\smallskip}\hline\noalign{\smallskip}
$\lambda$ (nm) & E$_{\rm low}$ (eV)  &\loggf  & Q  & $n_{\rm low}$ 
& $n_{\rm up}$ \\
\noalign{\smallskip}\hline\noalign{\smallskip}
 744.229  & 10.330 & -0.385 &  B+ & 4 & 10\\
 746.831  & 10.336 & -0.190 &  B+ & 4 & 10\\
 821.634  & 10.336 & +0.132 &  B+ & 4 &  9\\
 822.314  & 10.330 & -0.271 &  B+ & 4 &  9\\
 868.340  & 10.330 & +0.087 &  B+ & 4 &  8\\
 871.883  & 10.336 & -0.336 &  B+ & 4 &  8\\
1010.513  & 11.750 & +0.219 &  B+ & 8 & 17\\
1011.248  & 11.758 & +0.607 &  B+ & 8 & 17\\
1011.464  & 11.764 & +0.768 &  B+ & 8 & 17\\
1050.700  & 11.840 & +0.094 &  B  & 9 & 20\\
1052.058  & 11.840 & +0.010 &  B  & 9 & 20\\
1053.957  & 11.844 & +0.503 &  B  & 9 & 20\\
\noalign{\smallskip}\hline\noalign{\smallskip}
\end{tabular}
\end{center}
Notes: 
Error of $\log gf \le 0.03$~dex (Q=B+), $\le 0.08$~dex (Q=B);
$n_{\rm low}$ and $n_{\rm up}$ identify the level of the nitrogen
model atom used for the NLTE computations (cf.\ Fig.\,\ref{dep}).
\end{table}


\section {Models and line-formation codes}

\subsection{3D hydrodynamical model atmosphere}
Our analysis is based mainly on a 3-dimensional
radiation-hydrodynamics simulation (hereafter
3D model) computed with the \cobold\
code \citep{Freytag2002AN....323..213F,
Freytag2003CO5BOLD-Manual,Wedemeyer2004A&A...414.1121W}.
Some basic information about the setup of this numerical simulation 
can be found in \citet{oxy}, who used the same 
model to determine the solar oxygen abundance.
A full description of the model, including a critical review of its
performance in reproducing various observational constraints,  
will be given in \citet{solarmodels}.

\subsection{1D reference models}
For comparison, we also used several 1D solar models:
\begin{enumerate}
\item
The semi-empirical Holweger-M\"uller model (\citealt{hhsunmod, hmsunmod}, 
hereafter HM);
when necessary, this was placed on a mass column-density
scale, using
the opacity used to produce the \cobold\ model.
\item
A 1D model computed with the \LHD\ code, which
uses the same microphysics as \cobold\
and treats convection with the mixing-length
approximation, adopting \mlp=1.0 (see \citealt{zolfito} 
for further details), and
two further \xx\ models with \mlp\ of 0.5 and 2.0,
to investigate the dependence of the abundance on \mlp.
\item
The ATLAS9 solar model with the abundances of \citet{sunabboasp} as computed
by F. Castelli\footnote{http://wwwuser.oats.inaf.it/castelli/sun/ap00t5777g44377k1asp.dat}.
\item
A 1D model obtained by temporal and horizontal averaging over surfaces of 
equal (Rosseland) optical depth and over all snapshots.
We refer to this averaged model as \mD. 
\end{enumerate}

\subsection{Line-formation calculations}
The 3D spectrum synthesis computations are all
performed with 
\linfor\footnote{http://www.aip.de/$\sim$mst/Linfor3D/linfor\_3D\_manual.pdf},
which can also compute line formation using as input different type of 1D models.
For comparison in the case of 1D models, we also used the SYNTHE code
\citep{1993KurCD..18.....K,2005MSAIS...8...14K} in its Linux version 
\citep{2004MSAIS...5...93S,2005MSAIS...8...61S} in calculating 
synthetic spectra.
The advantage of SYNTHE over \linfor\ is that it can easily treat
hundreds of thousands of lines, while the
present version of \linfor\ is limited to a few tens of lines.


\section{Observational data}

Investigations of solar abundances found  in the literature
usually provide detailed discussions about both
the accuracy of the adopted model atmospheres and
the related uncertainties in the derived  abundances. 
However, the accuracy of the observational data is often
not taken into consideration, since, typically, solar abundance determinations rely
on a single observed spectrum.

\subsection{Differences between high quality solar spectra}

\begin{figure}
\resizebox{\hsize}{!}{\includegraphics[clip=true,angle=0]{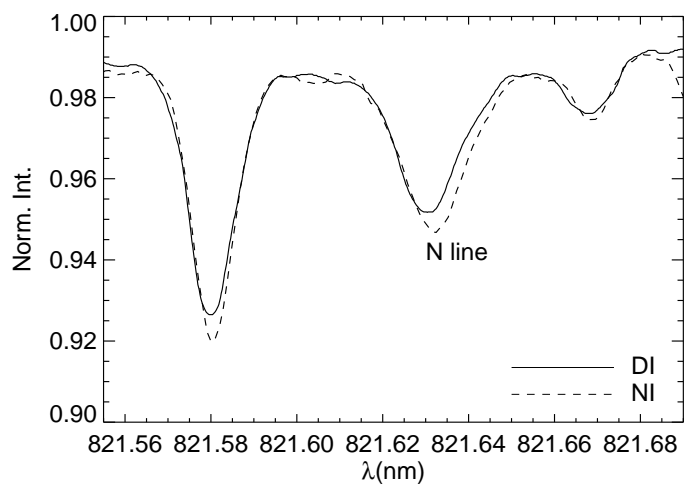}}
\caption{Observed profile of the \ion{N}{i} line at $\lambda$\, 821.6\,nm 
as extracted from the Delbouille (solid) and Neckel (dashed) solar 
disc-centre intensity spectral atlases. The reason for the considerable 
differences in unknown.}
\label{diff_n8216}
\end{figure}

In our study of the solar oxygen abundance \citep{oxy},
we noted that different high quality solar spectra
exhibit differences that were larger than expected
from their high S/N ratios.
To investigate the problem related to the differences between the solar
atlases in the case of nitrogen, we checked four selected lines. From
this non-exhaustive analysis, we conclude, as we did for other
elements, that for some, but not all, lines, the solar atlases
disagree.  This is the case for one of four nitrogen lines that
we considered as illustrated in Fig.\,\ref{diff_n8216}.
The presence of telluric absorption might explain
this difference. However, by analysing spectra of rapidly 
rotating hot stars,
there is no evidence of any telluric contamination, 
but the day-time sky can exhibit
absorption that is undetectable at night-time. An indisputable reason for the
differences still needs to be identified.
We strongly believe that
the astronomical community needs a new  high-quality solar atlas.

\subsection{Comparison of different EW measurements}

Inspection of the literature often reveals 
that even the measurement of equivalent widths (EWs) from one
and the same observed spectrum is less straightforward
than what could be expected. Different authors
use different approaches and measure different equivalent widths.
A recent example is the comparison of our EWs for oxygen
\citep{oxy} with those of \citet{asplund04}.
A long-running debate about the solar iron 
abundance has in part been  driven by
differences in the EWs \citep[see for example][and references therein]{HKB}.
For the nitrogen, the EW measurements differ also
between different works. 
Comparing the measurements of the EWs of two authoritative works 
such as \citet{grevesse} and \citet{biemont}, we conclude that
there is good agreement for our selected lines of nitrogen.
The authors measure similar values 
without any systematic differences being evident.

\subsection{Data used in this study}
In our analysis, we used the EWs of \citet{grevesse}
and \citet{biemont}, which are in excellent agreement.

For a few selected lines, we derived in addition the 
nitrogen abundance from detailed line profile fitting
(see Fig.\,\ref{FIT_HM}).
For this purpose, we used the two centre-disc intensity atlases 
(the ``Delbouille'' atlas, i.e., \citealt{delbouille},
\citealt{delbouilleir}, and the 
``Neckel'' intensity atlas, \citealt{neckelobs}),
as well as the two solar flux atlases (the ``Kurucz'' 
solar flux atlas \citealt{kuruczflux} and the ``Neckel''
solar flux atlas \citealt{neckelobs}). 
Further details about these atlases can be found
in the aforementioned references and \citet{oxy}.


\section{NLTE computations}

The nitrogen lines considered in our analysis are weak and form deep inside
the solar photosphere (see Fig.\,\ref{ttau_conft}, showing the 
EW contribution function for one of the nitrogen lines). From these 
facts, we do not expect that these lines are very sensitive to 
departures from Local Thermodynamic Equilibrium (LTE).
In her analysis of the statistical equilibrium of nitrogen
in the Sun, \citet{inga96} inferred that the deviations from 
LTE are small and negative, on average about $-0.05$\,dex.
In this paper, we confirm this result.

\subsection{1D NLTE corrections}
Since the nitrogen lines form at the same photospheric 
depth range in
both 3D and 1D (\mD, \xx) models (see 
contribution functions in Fig.\,\ref{ttau_conft}), this infers that
3D granulation effects do not play a fundamental role.
Since the contribution functions for 3D and 1D models
are similar, it follows that the computed EWs are also similar,
implying that the 3D corrections are small.
Without a code capable of solving the full 3D-NLTE problem for
nitrogen, we resort to 1D-NLTE calculations.

\begin{figure}
\resizebox{\hsize}{!}{\includegraphics[clip=true,angle=0]{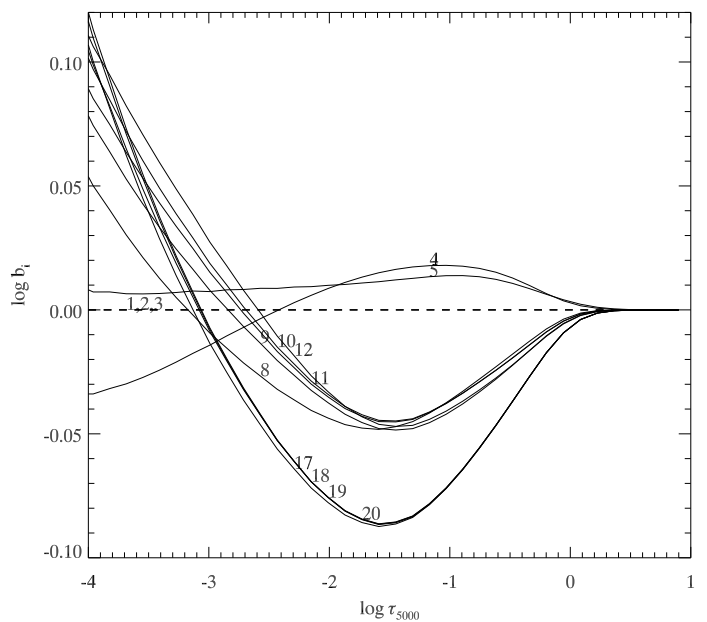}}
\caption{Departure coefficients $\log b_i = \log (n_{\rm NLTE}/n_{\rm LTE})$ of 
selected levels of neutral nitrogen as a function of optical depth $\log\tau_{5000}$. 
For each line involved in our abundance analysis, the identification of the 
upper and lower level is listed in Table\,\ref{atomicdata}. The shown $b_i$ refer to
the \mD\ model; similar results are obtained for the HM~model. 
}
\label{dep}
\end{figure}

We computed the departures from LTE for the \mD\ and HM models,
using the Kiel code \citep{SH}, with the model atom of \citet{inga96}.
The Kiel code uses a generalisation of the \citet{Drawin} formalism
to take into account excitation and ionisation of the nitrogen atoms by 
inelastic collisions with neutral hydrogen atoms. 
A scaling factor, ${\rm S_H}$, permits to reduce the efficiency of 
these collisions (${\rm S_H}<1$), to switch them off (${\rm S_H}=0$),
or to consider them ``in toto'' (${\rm S_H}=1$).
Consistent with the Kiel group and in line with \citet{inga96},
we favour ${\rm S_H}=1/3$, even though we computed
NLTE corrections in addition for ${\rm S_H}=0$ and ${\rm S_H}=1$.
In Fig.\,\ref{dep}, the departure coefficients of the atomic
levels relevant to the selected lines (see Table\,\ref{atomicdata})
are shown for the \mD\ model.

The NLTE corrections obtained from the \mD\ and the HM model are 
given in Table\,\ref{ansubselnlte}. They are added to the abundances 
obtained from the LTE analysis based on the 3D and the HM model, 
respectively, to obtain the final NLTE nitrogen abundances.

\subsection{Influence of horizontal temperature fluctuations}

Using the \mD\ model to compute departures from LTE is an extreme case in
which all horizontal temperature fluctuations are neglected. To estimate how
this approximation affects departures from LTE, we produced horizontal 
and time-averaged models by grouping columns into twelve bins according to their 
emergent continuum intensity in the vertical direction. A similar procedure was used 
by \citet{aufdenberg05} to estimate the effects of horizontal, temperature
inhomogeneities. 

\begin{figure}
\resizebox{\hsize}{!}{\includegraphics[clip=true,angle=0]{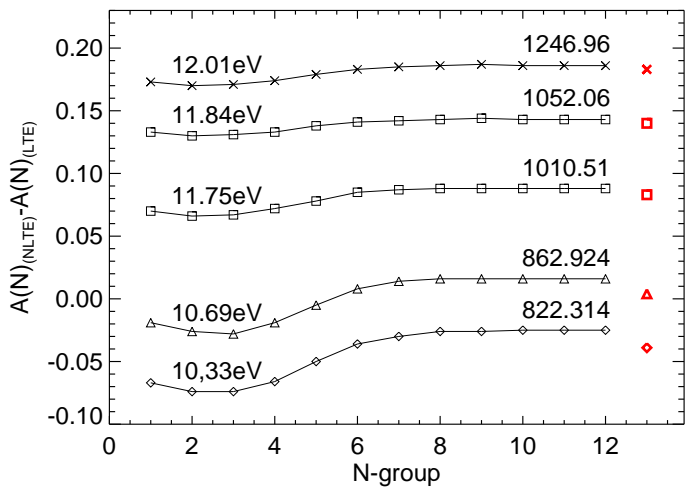}}
\caption{For five representative lines the 1D-NLTE corrections
from the twelve group-averaged models, ordered according to increasing
continuum intensity (radiation temperature) 
from left to right, are shown together with the
result for the global \mD\ model (rightmost, bold symbol).
The lowest curve is plotted at the true ordinate level, while each 
of the others curves is shifted up by 0.05\,dex with respect
to the previous one for clarity.}
\label{grpmod}
\end{figure}

These twelve components represent the different horizontal temperature structures
associated with the presence of granules and intergranular lanes.
Each component
has a weight that depends on its surface area fraction. The cool downdrafts are
associated with the groups of lowest intensity, and the warm upflows with those of
the highest intensity.  We computed NLTE corrections using each of these
components, which were treated as standard plane-parallel model atmospheres (1.5D approximation).
The components allowed us to study the variations in the NLTE corrections as a function 
of the temperature structure of the flow where the line is formed. The results,
for a selection of lines, are shown in Fig.\,\ref{grpmod}. 

Lines with
excitation energies higher than 11\pun{eV} exhibit little differences in
departures from LTE for the twelve group-averaged models. For the lines 
with lower excitation energies (E$_{\rm low} \approx 10.3$\pun{eV}),
the groups corresponding to the cool downdrafts (groups 1 to 6) have a 
correction that is distinctly larger (in absolute value) than for the 
groups corresponding to the warm upflows (groups 7 to 12). We consider 
the variation in the NLTE corrections of the various components, relative
to the result of the \mD\ model, as an indication of the uncertainty 
associated with the 1D approximation.

The variations in the NLTE-corrections are small, even
for the lower excitation lines, for which the difference between the extreme
groups is less than 0.05~dex. We therefore
expect that the results we obtained using the \mD\ model will
not disagree by more than 0.03~dex with the
results of a full 3D-NLTE computation.

\begin{table*} 
\caption{A(N)$_{\rm LTE}$ from selected \ion{N}{i} lines with EWs from the literature using \linfor}
\label{ansubsel}
\begin{center}
\begin{tabular}{rccccccccccccc}
\noalign{\smallskip}\hline\noalign{\smallskip}
$\lambda$ &\multicolumn{2}{c}{EW}&\multicolumn{2}{c}{A(N)3D}&\multicolumn{2}{c}{A(N)\mD}&\multicolumn{2}{c}{A(N)\xx}&\multicolumn{2}{c}{A(N)$_{\rm HM}$}& 3D-\mD & 3D-\xx(\mlp=1.0) \\
(nm) & \multicolumn{2}{c}{(pm)} & & & & & & & & & dex & dex\\
        & G    & B    & G     & B     & G     & B     & G     & B     & G     & B     &        &        \\
 (1)    &  (2) & (3)  & (4)   & (5)   & (6)   & (7)   & (8)   & (9)   & (10)  & (11)  & (12)   & (13)\\
\noalign{\smallskip}\hline\noalign{\smallskip}
  744.2 & 0.26 & 0.27 & 7.808 & 7.826 & 7.847 & 7.865 & 7.810 & 7.828 & 7.911 & 7.928 & $-0.039$ & $-0.002$ \\
  746.8 & 0.52 & 0.49 & 7.961 & 7.931 & 7.994 & 7.964 & 7.954 & 7.924 & 8.057 & 8.027 & $-0.033$ & $+0.008$ \\
  821.6 & 0.86 & 0.87 & 7.854 & 7.860 & 7.892 & 7.899 & 7.847 & 7.853 & 7.957 & 7.963 & $-0.039$ & $+0.007$ \\
  822.3 & 0.24 &      & 7.593 &       & 7.648 &       & 7.611 &       & 7.718 &       & $-0.055$ & $-0.018$ \\
  868.3 & 0.78 & 0.81 & 7.828 & 7.849 & 7.865 & 7.885 & 7.821 & 7.841 & 7.929 & 7.949 & $-0.037$ & $+0.007$ \\
  871.8 & 0.42 & 0.43 & 7.927 & 7.939 & 7.971 & 7.983 & 7.933 & 7.944 & 8.036 & 8.047 & $-0.044$ & $-0.006$ \\
 1010.5 & 0.18 &      & 7.956 &       & 8.022 &       & 7.976 &       & 8.098 &       & $-0.066$ & $-0.020$ \\
 1011.2 & 0.35 & 0.36 & 7.897 & 7.912 & 7.958 & 7.971 & 7.908 & 7.922 & 8.033 & 8.046 & $-0.060$ & $-0.011$ \\
 1011.4 & 0.55 & 0.54 & 7.976 & 7.966 & 8.028 & 8.019 & 7.976 & 7.967 & 8.101 & 8.092 & $-0.053$ & $-0.001$ \\
 1050.7 & 0.14 &      & 8.002 &       & 8.066 &       & 8.022 &       & 8.143 &       & $-0.064$ & $-0.020$ \\
 1052.0 & 0.08 &      & 7.829 &       & 7.896 &       & 7.853 &       & 7.975 &       & $-0.067$ & $-0.024$ \\
 1053.9 & 0.32 &      & 7.989 &       & 8.046 &       & 7.999 &       & 8.121 &       & $-0.057$ & $-0.010$ \\
\noalign{\smallskip}\hline\noalign{\smallskip}
 average&      &      & 7.885 & 7.890 & 7.936 & 7.941 & 7.892 & 7.897 & 8.007 & 8.007 & $-0.051$ & $-0.008$ \\
\noalign{\smallskip}\hline\noalign{\smallskip}
\end{tabular}
\end{center}
Notes:
Col.s with G are from \citet{grevesse}, with B are from \citet{biemont}.
Col.~(1) is the wavelength; col.s~(2)-(3) the EWs; col.s~(4)-(11) A(N) from 3D, 
\mD, \xx, and HM model, col.s~(12)-(13) two different 3D corrections.
\end{table*}


\section{Nitrogen abundance determinations}  

\subsection{LTE abundance from selected lines}

The following abundance analysis is based on a subsample of lines given 
in Table\,\ref{atomicdata}.
We determined individual LTE nitrogen 
abundances, A(N)$_{\rm LTE}$ using the 3D, \mD, \xx, and HM model,
respectively,  and
the equivalent width measurements by \citet{grevesse} and 
\citet{biemont} where available. The results are given in Table\,\ref{ansubsel}. 
In this Table and in the others following, we indicate
the abundances to three decimal places, following the 
prescription of \citet{bevington} of retaining one more
significant digit than dictated by the error.
For nitrogen lines, which are formed
in the deep photosphere, 
the 3D correction, defined as A(N)$_{\rm 3D}$--A(N)$_{\xx}$, column (13), 
is small, in the range 
$-0.024 ~ - ~ 0.008$\,dex.
The ``granulation abundance correction'', A(N)$_{\rm 3D}$--A(N)$_{\mD}$,
(indicated in Col.~(12) of Table\,\ref{ansubsel}), which measures  
the influence of horizontal fluctuations, is somewhat
larger, and negative for all the lines. 
We reported similar 
behaviour for the \ion{O}{i} 615.8\pun{nm} line in previous work (see \citealt{oxy}).

\subsection{Line-profile fitting}

We selected four clean and not too weak lines from the set of 
lines in Table\,\ref{atomicdata}
(746.8\pun{nm}, 821.6\pun{nm}, 868.3\pun{nm}, and 1011.4\pun{nm})
for which we fitted the line profile for all the four
observed atlases (two fluxes and two centre-disc
intensities), using 1D models with SYNTHE as a line formation code.
We discarded the line at 1011.4\pun{nm} for which we could 
not obtain a reliable fit; the abundance that we derived from this
line was also too high to be consistent with those found for other lines, 
although we note that the four different spectra consistently indicated
the same (high) abundance.
The line list used in the profile fitting of this line appears to be
incomplete. Nevertheless, we decide to retain this line in the abundance determination 
inferred from the EWs, because there it is consistent with the other values, indicating
that the line decomposition method used to determine EW is reliable. 

The A(N) derived from the line at 821.6\pun{nm} in the four different 
spectra exhibits a scatter of 0.050\,dex, 
when considering the \mD\,-NLTE abundance,
which is considerably larger than the scatter found for the other two lines.
The difference between the A(N) derived from the two disc-integrated spectra 
is about 0.06\,dex, while the A(N) from the two centre-disc spectra differ by 
about 0.13\,dex. 
Neither the two disc-integrated 
spectra nor the two centre-disc spectra do agree (see
Fig.\,\ref{diff_n8216}). The difference in EW is of the order of 15\%.
This is much larger than expected from the high S/N ratio
of the spectra.

\begin{figure}
\resizebox{\hsize}{!}{\includegraphics[clip=true,angle=0]{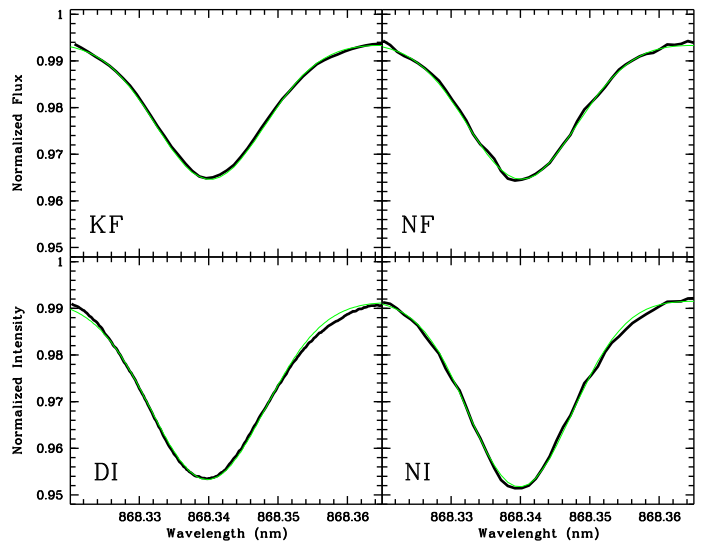}}
\caption{Results of the line profile fits for the 868.3\,nm line. 
The synthetic spectra computed with SYNTHE and using the Holweger-M\"uller 
model (thin lines) are superimposed on the observed
spectra from the four solar atlases (KF = Kurucz flux, NF = Neckel Flux,
DI = Delbouille Intensity, NI = Neckel Intensity; thick lines)
}
\label{FIT_HM}
\end{figure}

The line fitting results are given in Table\,\ref{tabfit}.
An example of the excellent agreement achieved in the line fitting 
using the HM model, is shown in Fig.\,\ref{FIT_HM}. 

The differences between the nitrogen abundance from both the fitting 
procedure and by matching the measured EWs can be explained by 
the influence of blending components, which are accounted for in
different ways in the line-fitting approach and in the EW method, respectively.
The different codes used may also be responsible for part of the difference.

\begin{figure}
\resizebox{\hsize}{!}{\includegraphics[clip=true,angle=0]{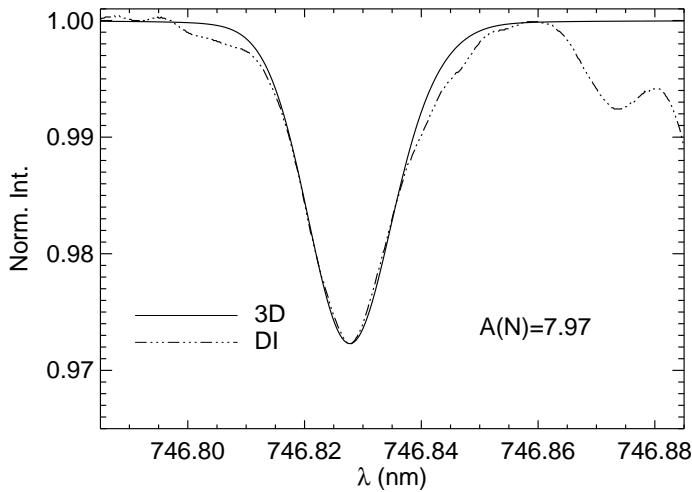}}
\caption{The unblended 3D profile of \ion{N}{i} $746.8$~nm is compared to 
the observed solar centre-disc profile of the Neckel Intensity atlas.
}
\label{FIT_3D}
\end{figure}

Since the majority of the nitrogen lines are blended, weak, or both, 
the line-fitting approach is, in principle, more appropriate.
However, line fitting with 3D spectra can become a 
tedious task, and we
therefore prefer to rely on equivalent width measurements.
We believe that we cannot achieve superior measurement
of the EWs than the careful works by \citet{grevesse} and \citet{biemont}. 
In principle, we could also measure the EWs in data from the other atlases,
but our above line fitting exercise implies that this would introduce
an additional scatter of the order of at most approximately 0.03\,dex, due
to the differences between atlases. As we shall see, this is much
less than the scatter derived from the different lines.
Our final choice is therefore to derive
A(N) from the EWs from \citet{grevesse} and \citet{biemont}, using
both 3D and 1D models.

As a consistency check, we also compared the 3D profiles with the observed 
spectra, as shown for one example in
Fig.\,\ref{FIT_3D}. The agreement is in general good, given that the blending 
line visible in the observed spectrum were not considered in the synthetic
3D profile.

\begin{table} 
\caption{A(N) of \ion{N}{i} from line fitting using 1D models with SYNTHE}
\label{tabfit}
\begin{center}
\begin{tabular}{cccccc}
\noalign{\smallskip}\hline\noalign{\smallskip}
Observed & $\lambda$&\multicolumn{3}{c}{A(N)$_{\rm LTE}$} & A(N)$_{\rm NLTE}$ \\
Spectrum &  (nm)      & HM & ATLAS9 & \mD\ & \mD\ \\
\noalign{\smallskip}\hline\noalign{\smallskip}
KF &  746.8 & 7.861 & 7.824 & 7.809 & 7.750 \\
NF &  746.8 & 7.874 & 7.837 & 7.827 & 7.768 \\
NI &  746.8 & 7.890 & 7.834 & 7.823 & 7.785 \\
DI &  746.8 & 7.908 & 7.849 & 7.840 & 7.802 \\
\noalign{\smallskip}\hline\noalign{\smallskip}
KF &  821.6 & 7.847 & 7.811 & 7.805 & 7.730 \\
NF &  821.6 & 7.918 & 7.870 & 7.867 & 7.792 \\
NI &  821.6 & 7.944 & 7.879 & 7.873 & 7.821 \\
DI &  821.6 & 7.808 & 7.752 & 7.746 & 7.694 \\
\noalign{\smallskip}\hline\noalign{\smallskip}
KF &  868.3 & 7.910 & 7.889 & 7.857 & 7.783 \\
NF &  868.3 & 7.906 & 7.877 & 7.857 & 7.783 \\
NI &  868.3 & 7.933 & 7.878 & 7.875 & 7.831 \\
DI &  868.3 & 7.932 & 7.886 & 7.882 & 7.838 \\
\noalign{\smallskip}\hline\noalign{\smallskip}
\end{tabular}
\end{center}
Notes:
Col.~(1) is the observed spectrum identification, KF: Kurucz Flux, NF: Neckel Flux,
NI: Neckel Intensity, and DI: Delbouille Intensity;
col.~(2) is the wavelength; col.s~(3)-(5) the A(N)$_{\rm LTE}$ from line fitting with
HM, ATLAS9, and \mD\  model;
col.~(6) is the A(N)$_{\rm NLTE}$ for the \mD\ model, obtained by adding
the NLTE-correction computed with $\mathrm{S_H}=1/3$ to the abundances given
in column (5).
\end{table}
%
\begin{table*}
\caption{A(N)$_{\rm LTE}$ and $\Delta$A(N)$_{\rm NLTE}$ from selected \ion{N}{i} lines with
disc-centre EWs from the literature obtained with \linfor\ and Kiel code for NLTE computations.}
\label{ansubselnlte}
\begin{center}
\begin{tabular}{rcccccccccc}
\noalign{\smallskip}\hline\noalign{\smallskip}
$\lambda$ (nm) &\multicolumn{2}{c}{A(N)3D$_{\rm LTE}$} & 
\multicolumn{3}{c}{$\Delta$A(N)$_{\rm NLTE}$~\mD}&
\multicolumn{2}{c}{A(N)HM$_{\rm LTE}$}&\multicolumn{3}{c}{$\Delta$A(N)$_{\rm NLTE}$~HM}\\
 &  & & \multicolumn{3}{c}{${\rm S_H}$} & & & \multicolumn{3}{c}{${\rm S_H}$}\\
        & G    & B  & 1/3 & 1 & 0 & G & B & 1/3 & 1  & 0  \\
 (1)    &  (2) & (3)  & (4)   & (5)   & (6)   & (7)   & (8)   & (9)   & (10)  & (11)\\
\noalign{\smallskip}\hline\noalign{\smallskip}
  744.2 & 7.808 & 7.826 & --0.034 & --0.026 & --0.038 & 7.911 & 7.928 & --0.038 & --0.028 & --0.042\\
  746.8 & 7.961 & 7.931 & --0.038 & --0.030 & --0.042 & 8.056 & 8.027 & --0.043 & --0.032 & --0.047\\
  821.6 & 7.854 & 7.860 & --0.052 & --0.037 & --0.064 & 7.957 & 7.963 & --0.057 & --0.041 & --0.071\\
  822.3 & 7.593 &       & --0.039 & --0.028 & --0.049 & 7.718 &       & --0.045 & --0.032 & --0.056\\
  868.3 & 7.828 & 7.849 & --0.047 & --0.034 & --0.061 & 7.929 & 7.949 & --0.051 & --0.036 & --0.065\\
  871.8 & 7.927 & 7.939 & --0.040 & --0.029 & --0.052 & 8.036 & 8.047 & --0.044 & --0.032 & --0.058\\
 1010.4 & 7.956 &       & --0.017 & --0.012 & --0.025 & 8.098 &       & --0.020 & --0.015 & --0.029\\
 1011.2 & 7.897 & 7.912 & --0.019 & --0.014 & --0.028 & 8.033 & 8.046 & --0.023 & --0.016 & --0.032\\
 1011.4 & 7.976 & 7.966 & --0.021 & --0.015 & --0.039 & 8.101 & 8.092 & --0.024 & --0.017 & --0.035\\
 1050.7 & 8.002 &       & --0.010 & --0.007 & --0.016 & 8.143 &       & --0.012 & --0.009 & --0.020\\
 1052.0 & 7.829 &       & --0.010 & --0.007 & --0.017 & 7.975 &       & --0.011 & --0.008 & --0.018\\
 1053.9 & 7.989 &       & --0.011 & --0.009 & --0.018 & 8.121 &       & --0.013 & --0.009 & --0.022\\
\noalign{\smallskip}\hline\noalign{\smallskip}
 average& 7.885 & 7.890 & --0.028 & --0.020 & --0.037 & 8.007 & 8.007 & --0.032 & --0.023 & --0.041\\
\noalign{\smallskip}\hline\noalign{\smallskip}
\end{tabular}
\end{center}
Notes:
Col.s with G are from \citet{grevesse}, with B are from \citet{biemont}.
Col.~(1) is the wavelength; col.s~(2)-(3) A(N) from 3D model,
col.s~(4)-(6) the corresponding NLTE corrections for ${\rm S_H}=1/3,~1,~0$, respectively;
col.s~(7)-(8) A(N) from HM model;
col.s~(9)-(11) the corresponding NLTE corrections for ${\rm S_H}=1/3,~1,~0$, respectively.
\end{table*}


\subsection{3D-NLTE abundance}

Table\,\ref{ansubselnlte} provides the NLTE corrections 
for both the \mD\ and the HM models. Our final 3D-NLTE nitrogen
abundance is obtained by averaging the 12 individual 3D-NLTE 
abundances (Col.~(2)+Col.~(4)), with equal weight.
The individual as well as the averaged NLTE-corrections are 
given in Col.s~(4), (5), and (6), 
respectively. The result is:
\begin{equation}
\begin{array}{c @{\rm A(N)=~} c @{~~~~~{\rm for} ~~~~~ {\rm S_H} =} c }
     & 7.85\pm 0.12 & 0    \\
     & 7.86\pm 0.12 & 1/3  \\
     & 7.87\pm 0.12 & 1    \\
\end{array}
\end{equation}
With the EWs from \citet{biemont}, 
A(N) remains the same while the scatter is reduced to 0.06\,dex.
Since the scatter is insensitive to the choice of ${\rm S_H}$
due to these lines being formed at similar depths in the
solar photosphere and having similar, and, in any case, small, NLTE corrections.

\section{Discussion}

\subsection{3D effects and NLTE-corrections}
A decisive point in favour of using 3D hydrodynamical models (or a
semi-empirical model) is that there is no need to invoke
mixing-length theory. All the nitrogen lines used in our
abundance determination arise from transitions between levels of high
excitation energy, so that they are formed deep in the solar
photosphere, in the range $-1<\log\tau<+1$, and hence are sensitive to
the choice of the mixing-length parameter in classical 1D models.
A change of $^{+1.0}_{-0.5}$ in the value of the mixing-length parameter
of 1.0 that we adopt,
produces a change in the abundance with the \xx\ model of on average
$^{-0.09}_{+0.05}$~dex.
The ``granulation abundance correction'',
A(N)$_{\rm 3D}$--A(N)$_{\mD}$, is always negative, $\approx -0.05$~dex
on average, in agreement with the results of \citet{stehol02}. 
The total 3D correction, defined as  A(N)$_{\rm 3D}$--A(N)$_{\xx}(\mlp=1.0)$ 
is only slightly negative for most of the considered lines.
Therefore, the 3D model provides a slightly  smaller abundance 
(by $\approx -0.01$~dex) than the \xx\ model.

The NLTE effects for the analysed lines are small and negative, so
that ${\rm A(N)_{\rm LTE} > A(N)_{\rm NLTE}}$. The average NLTE
correction, for ${\rm S_H}=1/3$, is about --0.03~dex.

 \begin{table} 
 \caption{Solar metallicity ($Z$ and $Z/X$) for different
 choices of C and Ne abundance \label{zeta}}
 \centering
 \begin{tabular}{ccccc}
 \hline\noalign{\smallskip}
  \multicolumn{2}{r}{A(C)} & 8.39& 8.52& 8.59\\
 \noalign{\smallskip}\hline\noalign{\smallskip}
 A(Ne)&\phantom{j}  & \multicolumn{3}{c}{ $Z$\,;\,$Z/X$}  \\
 \noalign{\smallskip}\hline\noalign{\smallskip}
 7.94 & & 0.0146\,;\,0.0199 & 0.0153\,;\,0.0209 & 0.0158\,;\,0.0216\\
 7.97 & & 0.0146\,;\,0.0199 & 0.0154\,;\,0.0210 & 0.0159\,;\,0.0218\\
 8.02 & & 0.0148\,;\,0.0202 & {\bf 0.0156\,;\,0.0213} & 0.0161\,;\,0.0220\\
 8.16 & & 0.0154\,;\,0.0210 & 0.0162\,;\,0.0221 & 0.0167\,;\,0.0228\\
 \noalign{\smallskip}\hline\noalign{\smallskip}
 \end{tabular}
 \end{table}

\subsection{Estimating the solar metallicity}

Our favoured values for the solar nitrogen abundance, A(N)=7.86, and
the solar oxygen abundance, A(O)=8.76 \citep{oxy}, imply a revision of
the solar metallicity $Z$.  The two other elements important to the determination of
the solar metallicity (in terms of fractional mass) are neon and carbon.

Neon is not measurable in the spectrum of the solar photosphere. The
usual procedure is to measure the ratio Ne/O in the solar corona, and
assume this value to be the same as in the photosphere.  We may
consider Ne/O=0.15 \citep{sunabboasp} or Ne/O=0.18 \citep{grevesse98},
implying A(Ne)=$7.94$ and A(Ne)=$8.02$, respectively.
By studying Ne abundances in nearby B-type stars \citet{morel08}
found a mean A(Ne)=$7.97\pm 0.07$, and suggest that this value should also be
representative of the solar Ne abundance. \citet{wang}
pointed out that the Ne/O ratios in \ion{H}{ii} regions and planetary
nebulae equal on average 0.25, and suggested that the current estimate
of the solar Ne/O ratio could be too low. If the solar Ne/O ratio was so large,
the Ne abundance would be $8.16$.  We may therefore
consider four possible values of A(Ne), ranging from $7.94$ to $8.16$.
For the carbon abundance, we have three choices, A(C)=$8.52$ according to
\citet{grevesse98}, A(C)=$8.39$ from \citet{sunabboasp}, or A(C)=$8.592$
from \citet{hhoxy}.  

The situation is summarised in Table\,\ref{zeta}, where our A(N) and A(O) 
are used with all possible combinations of A(Ne) and A(C)
to provide $Z$ and $Z/X$. The metallicity $Z$ spans the range $0.0146$ to 
$0.0167$, while $Z/X$ spans the range $0.0199$ to $0.0228$. All these 
values are considerably higher than the values recommended by 
\citet{sunabboasp} ($Z=0.0122$ and $Z/X=0.0165$), and the increase is 
mainly driven by our higher oxygen abundance, although the adopted values of N, C, 
and Ne also play a role.

\subsection{Metallicity from helioseismology}

This upward revision alleviates the tension between photospheric
abundances and helioseismic data, and the two may 
essentially agree. Using the
solar {\it relative} composition of \citet{sunabboasp} 
and OP opacities, \citet{basu} inferred from the
depth of the convection zone that $Z/X = 0.0218\pm 0.0008$, although the computation
should be repeated with our solar composition, but we note that this
value is already within the range of the $Z/X$ in Table\,\ref{zeta}.
Using the \citet{sunabboasp}
solar composition and one of their test models, \citet{chaplin} derived $Z=0.0161\pm
0.00008$ from low-degree modes, again in the range of values of 
Table\,\ref{zeta}.  Using the
more extensive set of models obtained from a Monte Carlo simulation,
\citet{chaplin} inferred values of between $Z=0.0187$ and $Z=0.0239$.  These values are
definitely higher than any of the values in Table\,\ref{zeta}.  From
the solar sound-speed profile, \citet{antia} derive $Z=0.0172\pm
0.002$. Again, this value is higher than any value in Table\,\ref{zeta},
even though it is within 2$\sigma$ of the highest value in the table.

\section{Conclusions}

Our recommended value of the solar nitrogen abundance is 
A(N)=$7.86\pm 0.12$, which takes into account 3D effects and deviations
from LTE. This value is between the value inferred by \citet{sunabboasp} 
($7.78\pm 0.06$) and that found by \citet{grevesse98} ($7.92\pm 0.06$).
Our A(N) is slightly lower, although consistent within errors,
with the result found by \citet[][A(N)=$7.931\pm 0.111$]{hhoxy}. 

In the course of our analysis, we found considerable
differences between the various solar spectral atlases,
introducing additional uncertainties in the  abundances
derived from individual lines. The disagreement is by
far larger than what can be attributed to the noise in the spectra.
This fact has been pointed out already in our previous solar abundance studies.
We believe that the quality of the available solar spectral atlases, 
which date back 30 or more years, should be improved by making use of up-to-date
technology.

Our most robust estimate of the solar $Z$ is obtained by adopting the abundances 
of \citet{grevesse98} for all elements except for oxygen, nitrogen, and neon,
for which we take our own abundances, A(O)=8.76, A(N)=7,86, and  A(Ne)=8.02.
The latter results from the ratio Ne/O=0.18 given by \citet{grevesse98}, 
combined with our oxygen abundance. In this way, the resulting solar metallicity 
is $Z=0.0156$ and $Z/X=0.0213$, respectively.

Pending a new determination of the solar C abundance, 
we can say that our new results for N and O represent a significant
step in forward reconciling 
photospheric abundances with helioseismic constraints on $Z$.

\begin{acknowledgements}
EC, HGL and PB acknowledge support from EU contract MEXT-CT-2004-014265
(CIFIST). The authors thank Katharina Lodders for her input on meteoritic
abundances.  E.M. is grateful to the Observatoire de Paris and to CNRS for a
research stage in Paris during the data analysis procedure of this work.
\end{acknowledgements}


\begin{thebibliography}{}
\bibitem[Anders \& Grevesse(1989)]{ag89} Anders, E., \& Grevesse, N.\ 1989, \gca, 53, 197 

\bibitem[Antia \& Basu(2006)]{antia} Antia, H.~M., \& Basu, S.\ 2006, \apj, 644, 1292 

\bibitem[Asplund et al.(2004)]{asplund04} Asplund, M., Grevesse, 
N., Sauval, A.~J., Allende Prieto, C., \& Kiselman, D.\ 2004, \aap, 417, 
751 

\bibitem[Asplund et al. (2005)]{sunabboasp} Asplund, M., Grevesse, 
N., \& Sauval, A.~J.\ 2005, ASP Conf.~Ser.~336: Cosmic Abundances as 
Records of Stellar Evolution and Nucleosynthesis, 336, 25 

\bibitem[Aufdenberg et al. (2005)]{aufdenberg05} Aufdenberg, J.~P., Ludwig,
  H.-G., Kervella, P. \ 2005, \apj, 633, 424 

\bibitem[Basu \& Antia(1997)]{BA97} Basu, S., \& Antia, H.~M.\ 1997, \mnras, 287, 189 


\bibitem[Basu \& Antia(2008)]{basu} Basu, S., \& Antia, H.~M.\ 2008, \physrep, 457, 217 


\bibitem[Basu et al.(2007)]{basu07} Basu, S., Chaplin, W.~J., 
Elsworth, Y., New, R., Serenelli, A.~M., 
\& Verner, G.~A.\ 2007, \apj, 655, 660 

\bibitem[Bevington \& Robinson(2003)]{bevington} Bevington, P.~R., \& Robinson, D.~K.\ 2003, 
Data reduction and error analysis for the physical sciences, 
3rd ed., by Philip R.~Bevington, and Keith D.~Robinson.~Boston, 
MA: McGraw-Hill, ISBN 0-07-247227-8, 2003


\bibitem[Bi{\'e}mont et al.(1990)]{biemont} Bi{\'e}mont, E., 
Froese Fischer, C., Godefroid, M., Vaeck, N., 
\& Hibbert, A.\ 1990, 3rd International 
Colloquium of the Royal Netherlands Academy of Arts and Sciences, 59

\bibitem[Caffau \& Ludwig(2007)]{zolfito} Caffau, E., \&
Ludwig, H.-G.\ 2007, \aap, 467, L11

\bibitem[Caffau et al.(2008)]{oxy} Caffau, E., Ludwig, 
H.-G., Steffen, M., Ayres, T.~R., Bonifacio, P., Cayrel, R., Freytag, B., 
\& Plez, B.\ 2008, A\&A, in press,  arXiv:0805.4398 

\bibitem[Chaplin et al.(2007)]{chaplin} Chaplin, W.~J., 
Serenelli, A.~M., Basu, S., Elsworth, Y., New, R., 
\& Verner, G.~A.\ 2007, \apj, 670, 872 

\bibitem[Delbouille et al.(1973)]{delbouille} Delbouille, L., 
Roland, G., \& Neven, L.\ 1973, Liege: Universite de Liege, Institut 
d'Astrophysique, 1973,  

\bibitem[Delbouille et al.(1981)]{delbouilleir}Delbouille L., Roland G., Brault, Testerman 1981;
``Photometric atlas of the solar spectrum from 1850 to 10,000 cm$^{-1}$'',
http://bass2000.obspm.fr/solar\_spect.php

\bibitem[Drawin (1969)]{Drawin} Drawin, H.W., 1969, Z. Physik 225, 483

\bibitem[{{Freytag} {et~al.}(2002){Freytag}, {Steffen}, \&
  {Dorch}}]{Freytag2002AN....323..213F}
{Freytag}, B., {Steffen}, M., \& {Dorch}, B. 2002, Astronomische Nachrichten,
  323, 213

\bibitem[{{Freytag} {et~al.}(2003){Freytag}, {Steffen}, {Wedemeyer-B{\"o}hm},
  \& {Ludwig}}]{Freytag2003CO5BOLD-Manual}
{Freytag}, B., {Steffen}, M., {Wedemeyer-B{\"o}hm}, S., \& {Ludwig}, H.-G.
  2003, {CO5BOLD User Manual},
  \verb|http://www.astro.uu.se/~bf/co5bold_main.html|

\bibitem[Grevesse et al.(1990)]{grevesse} Grevesse, N., Lambert, D.~L., 
Sauval, A.~J., van Dishoeck, E.~F., 
Farmer, C.~B., \& Norton, R.~H.\ 1990, \aap, 232, 225

\bibitem[Grevesse 
\& Sauval(1998)]{grevesse98} Grevesse, N., \& Sauval, A.~J.\ 1998, Space Science Reviews, 85, 161

\bibitem[Grevesse 
\& Sauval(2000)]{gs00} Grevesse, N., \& Sauval, A.~J.\ 2000, Origin of Elements in the Solar 
System, Implications of Post-1957 Observations, 261 

\bibitem[Holweger(1967)]{hhsunmod} Holweger, H.\ 1967, 
Zeitschrift fur Astrophysik, 65, 365 

\bibitem[Holweger \& M\"uller(1974)]{hmsunmod} Holweger, H., \& 
M\"uller, E.~A.\ 1974, \solphys, 39, 19 

\bibitem[Holweger et 
al.(1995)]{HKB} Holweger, H., Kock, M., \& Bard, A.\ 1995, \aap, 296, 233 

\bibitem[Holweger(2001)]{hhoxy} Holweger, H.\ 2001, AIP
Conf.~Proc.~598: Joint SOHO/ACE workshop ''Solar and Galactic 
Composition'', 598, 23


\bibitem[Houziaux(1961)]{Houziaux} Houziaux, L.\ 1961, 
Zeitschrift fur Astrophysik, 53, 237 

\bibitem[{{Kurucz}(1993b)}]{1993KurCD..18.....K}
{Kurucz}, R. 1993b, SYNTHE Spectrum Synthesis Programs and Line
Data.~Kurucz CD-ROM No.~18.~Cambridge, Mass.: Smithsonian Astrophysical
Observatory, 1993., 18

\bibitem[Kurucz(2005a)]{kuruczflux} Kurucz, R.~L.\ 2005a, Memorie 
della Societ\`a Astronomica Italiana Supplementi, 8, 189

\bibitem[{{Kurucz}(2005b)}]{2005MSAIS...8...14K}
{Kurucz}, R.~L. 2005b, Memorie della Societ\`a Astronomica
  Italiana Supplementi, 8, 14

\bibitem[Lambert(1978)]{lambert78} Lambert, D.~L.\ 1978, \mnras, 
182, 249 

\bibitem[Lambert \& Swings(1967)]{LS67} 
Lambert, D.~L., \& Swings, J.~P.\ 1967, The Observatory, 87, 113 

\bibitem[Ludwig et al.(2009)]{solarmodels} Ludwig, H.-G.,
Steffen, M., Freytag, B., Caffau, E., Bonifacio, P., \& Plez, B., 
2009 (\aap, in preparation)

\bibitem[Moore(1945)]{Moore} Moore, C.~E.\ 1945, 
Contributions from the Princeton University Observatory, 20, 1 

\bibitem[Moore et al.(1966)]{moore66} Moore, C.~E., Minnaert, 
M.~G.~J., 
\& Houtgast, J.\ 1966, National Bureau of 
Standards Monograph, Washington: 
US Government Printing Office (USGPO), 1966,  

\bibitem[Morel 
\& Butler(2008)]{morel08} Morel, T., \& Butler, K.\ 2008, ArXiv e-prints, 806, arXiv:0806.0491

\bibitem[Mugglestone(1958)]{Mugglestone} Mugglestone, D.\ 1958, 
\mnras, 118, 432 

\bibitem[Neckel \& Labs(1984)]{neckelobs} Neckel, H., \& Labs, 
D.\ 1984, \solphys, 90, 205 

\bibitem[Palme \& Jones(2003)]{palme05} Palme, H., \& Jones, A.\ 2003, 
Meteorites, Comets and Planets: Treatise on Geochemistry, Volume 1, 41

\bibitem[Ralchenko(2005)]{Ralchenko} Ralchenko, Y.\ 2005, Memorie 
della Societa Astronomica Italiana Supplement, 8, 96,
{http://physics.nist.gov/PhysRefData/ASD/index.html}

\bibitem[Rentzsch-Holm(1996)]{inga96} Rentzsch-Holm, I.\ 1996, \aap, 305, 275 

\bibitem[{{Sbordone}(2005)}]{2005MSAIS...8...61S}
{Sbordone}, L. 2005, Memorie della Societ\`a Astronomica Italiana Supplementi, 8, 61

\bibitem[{{Sbordone} {et~al.}(2004){Sbordone}, {Bonifacio}, {Castelli}, \&
  {Kurucz}}]{2004MSAIS...5...93S}
{Sbordone}, L., {Bonifacio}, P., {Castelli}, F., \& {Kurucz}, R.~L. 2004,
  Memorie della Societ\`a Astronomica Italiana Supplementi, 5, 93

\bibitem[Steenbock \& Holweger(1984)]{SH} Steenbock, W., 
\& Holweger, H.\ 1984, \aap, 130, 319

\bibitem[Steffen \& Holweger(2002)]{stehol02} Steffen, M., 
\& Holweger, H.\ 2002, \aap, 387, 258

\bibitem[Swensson(1967)]{Swensson} Swensson, J.~W.\ 1967, 
Zeitschrift fur Astrophysik, 66, 156 

\bibitem[Swensson et al.(1970)]{Swensson70} Swensson, J.~W., 
Benedict, W.~S., Delbouille, L., 
\& Roland, G.\ 1970, Memoires of the Societe Royale des Sciences de Liege, 5

\bibitem[Wang \& Liu(2008)]{wang} Wang, W., \& Liu, X.~-.\ 2008, MNRAS in press, arXiv:0806.2288 

\bibitem[{{Wedemeyer} {et~al.}(2004){Wedemeyer}, {Freytag}, {Steffen},
  {Ludwig}, \& {Holweger}}]{Wedemeyer2004A&A...414.1121W}
{Wedemeyer}, S., {Freytag}, B., {Steffen}, M., {Ludwig}, H.-G., \& {Holweger},
  H. 2004, \aap, 414, 1121



\end{thebibliography}
\end{document}